\documentclass[floatfix,nofootinbib,twocolumn,showpacs,aps,prd,10pt,showpacs]{revtex4-1}

\usepackage[colorlinks,citecolor=teal]{hyperref}
\usepackage{slashed}
\usepackage{amsmath}
\usepackage{amssymb}
\usepackage{graphicx}
\newcommand{\punto}{\!\cdot\!}
\newcommand{\bs}[1]{{\boldsymbol #1}}
\usepackage{xcolor}

\begin{document}

\title{Metastable pions in dense media}

\author{Marcelo Loewe$^{1,2,3}$, Alfredo Raya$^{4}$, Cristi\'an 
Villavicencio$^{5}$} 
\affiliation{$^1$Instituto de F\'isica, Pontificia Universidad Cat\'olica de Chile, Casilla 
306, Santiago 22, Chile\\
$^2$Centre for Theoretical and Mathematical Physics, and Department of Physics, University of Cape
Town, Rondebosch 7700, South Africa\\
$^3$Centro Cient\'{\i}fico Tecnol\'ogico de Valpara\'{\i}so -- CCTVAL, Universidad T\'ecnica Federico Santa Mar\'ia, Casilla 110-V, Valpara\'{\i}so, Chile\\
  $^4$Instituto de F\'isica y Matem\'aticas, Universidad Michoacana de San 
Nicol\'as de Hidalgo, Edificio C-3, Ciudad Universitaria, C.P. 58040, Morelia, 
Michoac\'an, Mexico\\
  $^5$Departamento de Ciencias B\'asicas, Facultad de Ciencias, Universidad del B\'io-B\'io, Casilla 447, Chill\'an, Chile.}

\begin{abstract}
We study the leptonic decay of charged pions  in a compact star environment. Considering leptons as a degenerated Fermi system, pions are tightly constrained to decay into these particles because their Fermi levels are occupied. Thus, pion decay is only possible through thermal fluctuations. Under these circumstances, pion lifetime is larger and hence can be considered to reach a metastable state. We explore restrictions under which such a metastability is possible. We also study conditions under which  pions and leptons already in chemical equilibrium can reach simultaneously the thermal equilibrium, and obtain the neutrino emissivity from metastable pions. Scenarios that favor this metastable state are protoneutron stars.
\end{abstract} 
\pacs{}

\maketitle

\section{Introduction}  
The study of pions  in a condensed state has widely been explored
in different contexts and frameworks~\cite{Baym:1973zk, Sawyer:1973fv, Campbell:1974qt,  Maxwell:1977zz,  Son:2000by, Loewe:2002tw, He:2005nk, Andersen:2006ys, Sun:2007fc, Nakazato:2008su, Gubina:2012wp, Ayala:2012dk, Loewe:2013coa, Colucci:2013zoa,Mammarella:2015pxa, Carignano:2016rvs, Ebert:2016hkd}. 
In particular, compact stars may provide a natural scenario for such a state of matter.
The possibility of  generating pions in a condensed state in compact stars, however, is still matter of discussion. 
For the formation of a charged-pion condensate in a chemically  equilibrated system, it is necessary that the electron chemical potential reaches values up to the pion mass. 
Nevertheless, at high enough baryon density, the formation of hyperons is also possible and this process tends to reduce the electron chemical potential~\cite{Balberg:1998ug, glendenning,report2000, Ohnishi:2008ng}. 
On the other hand, at the inner core, the formation of kaon condensate is favored in comparison with pion condensate due to a reduction of kaon mass \cite{Pons:2000xf, Pons:2000iy, Beane:2000ms, Muto:2001uv, Menezes:2005ic, Brown:2007ara, Yamamoto:2008zw}. Moreover, it is expected that a  pion-nucleon  $s$-wave repulsive  interaction  increases the pion mass although the $p$-wave attractive potential produces the opposite effect \cite{Ishizuka:2008gr, Ohnishi:2008ng}.
All in all, there are many dense nuclear effects affecting the pion mass and leptonic decay constant  and thus the range of values for these observables is broadened \cite{Meissner:2001gz, Holt:2014hma}.
Such considerations make it unclear to distinguish whether the electron chemical potential may reach the pion mass, but definitely it can reach values close to the condensation point.

A vast portion of the existing literature regarding compact stars does not incorporate pions in the equation of state (EOS) because these studies are  mainly based on mean field models. 
In the absence of a  pion condensate, dynamical pions do not play a significant role in such models because they immediately decay and mostly  behave as an interaction mediating particle \cite{glendenning}.

If a pion production mechanism exists, the pion density number is negligible at low temperature by exponential suppression.  
However, this is not the case in protoneutron stars, where temperature reaches values higher than  1~MeV ($\sim 10^{10}$~K).
Improvements on the study of pions in hot and dense media in heavy ion collision experiments at moderate energies have been conducted,  showing an important increase of  the $\pi^-/\pi^+$ ratio at high baryon density \cite{Xu:2009fj, Xiao:2013awa, Feng:2015vra}. 
Thus, an increase of the electron chemical potential is expected, particularly as density increases.
So, at least in protoneutron stars, an important asymmetric charged-pion production is expected.
However, due to the short pion decay time, it is  reasonable to ignore these pions in the EOS and only consider them as a sudden source for neutrinos in the cooling process.

The 99.99\% of the charged pions decay into muons and muonic neutrinos, and rarely into electrons and electronic neutrinos as a subleading channel. 
Since muons and electrons inside a compact star are degenerated, 
almost all states are occupied up to Fermi levels and the only way for a negative pion to decay into leptons is through thermal fluctuations. 
In other words, there is not enough phase space available for the byproduct of a charged-pion decay, which makes it an extremely slow process. 
This observation implies that pions are in a  metastable state \cite{Loewe:2011tm}.
Moreover, lepton decay rate may be induced through thermal fluctuations by considering scattering with neutrinos from the thermal bath, generating as a result an increase of the number of $\pi^-$ in the thermal bath. 
When both the pionic and leptonic decay rates become of the same order, both types of particles reach thermal equilibrium simultaneously \cite{Weldon:1983jn, Kuznetsova:2008jt, Kuznetsova:2010pi}. 

In this article we explore the issue of charged-pion metastability and conditions under which it would be possible. 
We also discuss some phenomenological consequences of such a state. For this purpose, this article is organized as follows: 
In Sec.~\ref{considerations} we present the initial considerations for the system we are dealing with. 
In Sec.~\ref{decay_width} we calculate the pion decay rate in dense lepton medium and explore the case when there is  chemical equilibrium between pions and leptons, and nonchemical equilibrium with pions in rest frame. 
In Sec.~\ref{metastability} we show the conditions for metastability and we define the critical parameters to reach such a state. 
We calculate the lepton decay rate in Sec.~\ref{pi-l_eq} and the conditions in parameter space where pion-lepton  thermal equilibrium can be reached  simultaneously. 
The neutrino emissivity through pion decay is calculated in Sec.~\ref{nu_emission} and compared with the URCA process emissivity. 
Finally we present our conclusions and future perspectives in Sec.~\ref{conclusions}.

\section{Initial considerations}
\label{considerations}

Metastability of negative charged pions in a degenerate-muon and -electron environment is due to Pauli blocking, which suppresses
the meson decay into leptons.
This condition is achieved by most of the existing neutron star models.
In this section we summarize the general considerations adopted, the models to be used, and the notation employed throughout the article.

\begin{itemize}

\item
Charged-pion number is conserved in normal phase, namely,
 $\mu_\pi\equiv\mu_{\pi^-} = -\mu_{\pi^+}$, and $\mu_\pi < m_\pi$.

\item
Charged leptons are degenerated, which means that we consider all lepton chemical potentials bigger than their respective masses, $\mu_\ell > m_\ell$, where $\ell$ stands for muon ($\mu$) or electron ($e$).
Masses considered here are $m_\mu=105.6$~MeV, $m_e=0.5$~MeV, and neutrinos ($\nu$) are considered  massless.

\item
The nuclear medium modifies hadronic parameters. 
For simplicity, and because we attempt to describe weak interaction effects only, we do not specify particular values for $m_\pi$ and $f_\pi$ since we do not know \emph{a priori} the influence of dense nuclear matter on these parameters.
However, we always consider $m_\pi>m_\ell$.

\end{itemize}

\bigskip

Weak interactions between pions and leptons are described by the effective Fermi model
\begin{equation}
{\cal L}_{\pi\ell} = 
f_\pi G_F[ 
\bar\psi_{\nu_\ell} \slashed{D}\pi^+(1-\gamma_5)\psi_{\ell}
+\bar\psi_{\ell}\slashed{D}\pi^-(1-\gamma_5)\psi_{\nu_\ell}],
\label{L_pi-l}
\end{equation}
where the derivative 
\begin{equation}
D_\alpha\pi^\mp=(\partial_\alpha\mp i\mu_\pi \delta_{\alpha 0})\pi^\mp
\end{equation}
includes the charged-pion chemical potential.
The value used for the Fermi constant is $G_F=1.17\times 10^{-5}$~GeV$^{-2}$.

The free Lagrangians for pions, leptons, and neutrinos are, respectively, 
\begin{align}
{\cal L}_{\pi} &= D_\alpha\pi^+ D^\alpha\pi^- - m_\pi^2\pi^+\pi^-,\\
{\cal L}_\ell &= \bar\psi_\ell [i\slashed{\partial}+\mu_\ell\gamma_0-m_\ell]\psi_\ell\\
{\cal L}_{\nu_\ell} &= \bar\psi_{\nu_\ell}[i\slashed{\partial}+\mu_{\nu_\ell}\gamma_0]\psi_{\nu_\ell}.
\label{L_f}
\end{align}

\bigskip

We define the particle four-momentum as $p$ for pions, $q$ for leptons and $k$ for neutrinos. 
The energy of pions, charged leptons and neutrinos is defined as 
\begin{equation}
E_\pi  = \sqrt{{\boldsymbol p}^2+m_\pi^2},
\quad
E_\ell = \sqrt{{\boldsymbol q}^2+m_\ell^2},
\quad
E_{\nu_\ell} = |{\boldsymbol k}|,
\end{equation}
respectively.
Also, we define the corresponding particle number density distributions as
\begin{align}
n_{\pi^-} &= n_B(E_\pi-\mu_\pi),\\
n_\ell &= n_F(E_\ell -\mu_\ell),\\
n_{\bar\nu_\ell} &= n_F(E_{\nu_\ell}+\mu_{\nu_\ell}),
\end{align}
where $n_B(x)=(e^{x/T}-1)^{-1}$ and $n_F(x)= (e^{x/T}+1)^{-1}$ represent the Bose-Einstein and  Fermi-Dirac distributions, respectively.

\bigskip

Our treatment includes thermal effects, 
which we consider within the Matsubara formalism.
For this purpose, we change the four-momenta zeroth components $l_0$ to the respective Matsubara frequencies at a given temperature $T$ according to  $p_0= i2n_p\pi T$ for pions, $q_0 = i(2n_q+1)\pi T$ for leptons, and $k_0= i(2n_k+1)\pi T$ for neutrinos, for $n_p,\ n_q,\ n_k \in \mathbb{Z}$. Correspondingly, the momentum integrals change to $\int dl_0 = i 2\pi T\sum_{n}$ for bosons and fermions. 

\bigskip

As we mentioned, nuclear interactions affect considerably pion dynamics. 
In particular, it is not well understood whether the repulsive $s$ wave or the attractive $p$ wave dominates \cite{Xu:2009fj}, so we consider the effective pion mass undetermined.
However, it can be observed in hadronic models  that medium effects are incorporated through the pion self-energy and the field renormalization strength~\cite{Meissner:2001gz, Goda:2013npa}. 
It is always possible to expand these quantities at low momentum and the result is a freelike particle propagator with modified mass, and pion velocity~\cite{Goda:2013npa}. Here, for simplicity, we consider only the pion mass modification.

In the same way, the pion decay constant can be obtained expanding at low momentum of the axial current using PCAC \cite{Meissner:2001gz, Goda:2013npa} or by adopting  the Brown-Rho scaling \cite{Holt:2014hma}. We adopt the latter scaling for $f_\pi$ as well as the nucleon effective masses in this work.

\section{In-medium pion decay rate}
\label{decay_width}

In order to find the conditions for metastability, let us first examine in detail the in-medium pion decay rate, which we derive using the optical theorem. 
The same results are obtained from the Fermi golden rule.

The charged-pion propagator must be understood as the creation of a pion with a given charge an the annihilation of a pion with the opposite charge.
Since there is isospin asymmetry, the propagation is different for $\pi^+$ and $\pi^-$.
For our purposes, we need to calculate the retarded charged-pion propagator in momentum space\footnote{
A decay rate cannot be defined in the remote past; that is the reason why the retarded propagator is used. 
Usually there is no difference at all in vacuum, but at finite temperature and density, the imaginary part of the retarded and the time-ordered self-energy  produce different results. 
}, including self-energy corrections generated through weak coupling with leptons, namely,
\begin{equation}
D_{\pi^\mp}^{\mathrm{ret}}(p) =\left.\frac{i}{(p_0\pm \mu_\pi)^2-E_\pi^2-\Pi(\pm p)}\right|_{p_0\to p_0+i\epsilon},
\end{equation}
where $\Pi(p)$ is the self-energy.

Neglecting the real part of the self-energy and expanding around the physical real pole $p_0=E_\pi-\mu_\pi$, the negative-pion propagator can be  written in the nonrelativistic Breit-Wigner form
\begin{equation}
D_{\pi^-}^{\mathrm{ret}}(p) \approx 
\frac{i}{(p_0+\mu_\pi)^2-E_\pi^2 +  iE_\pi \Gamma_{\pi^-}}\;,
\end{equation}
where the momentum dependent decay rate is defined as
\begin{equation}
\Gamma_{\pi^-} = -\frac{1}{E_\pi}\mathrm{Im} ~\Pi(E_\pi-\mu_\pi+i\epsilon,~{\boldsymbol p}).
\label{def-Gamma}
\end{equation}
At zero temperature, this quantity is just the decay width of the pion. 
Nevertheless, at finite temperature, the meaning of this quantity should be interpreted differently (see below).
 \begin{figure}
\includegraphics{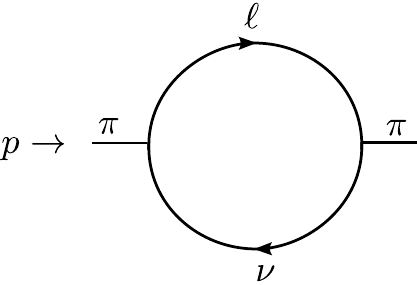}
\caption{Feynman diagram representing the pion self-energy from weak interactions in a thermal bath.}
\label{pionSE}
\end{figure}

We proceed to calculate the leptonic contribution to the pion self-energy. 
This process is shown diagrammatically in Fig.~\ref{pionSE} and can be written as
\begin{equation}
-i\Pi(p) = \mathrm{tr}\int \frac{d^4k}{(2\pi)^4}V(p) S_\ell(q) V(p) S_{\nu_\ell}(k)\;,
\end{equation}
where $S_\ell$ and $S_{\nu_\ell}$ are the time-ordered lepton and neutrino propagators, $q=p+k$ as momentum conservation requires, and the vertex obtained from the interaction term in Eq. (\ref{L_pi-l})
is defined as
\begin{align}
V(p) &=  G_F f_\pi (\slashed{p}+\mu_\pi\gamma_0)(1-\gamma_5).
\label{vertex}
\end{align}
Performing the traces over Dirac indices and summing over the Matsubara frequencies using the techniques described in \cite{lebellac}, we set the retarded condition $p_0\to p_0+i\epsilon$ and take the imaginary part of the self-energy. 
The general form for $\Gamma_{\pi^-}$ is 
\begin{multline}
\Gamma_{\pi^-} =  
\int \!\! \frac{d^3k}{(2\pi)^3}\sum_{s,t=\pm 1}
\frac{-2\pi G_F^2f_\pi^2}{ sE_\ell\, tE_{\nu_\ell}\,E_\pi }
\Big\{m_\pi^2[s E_\ell \, tE_{\nu_\ell}+{\boldsymbol q}\punto{\boldsymbol k}]
\\ \shoveright{
-2[ s  E_\ell\, E_\pi -{\boldsymbol p}\punto {\boldsymbol q}]
[ tE_{\nu_\ell}\, E_\pi  +{\boldsymbol p}\punto{\boldsymbol k}]
\Big\}}
\\ \shoveright{
\left[1-n_F(sE_\ell-\mu_\ell)-n_F(tE_{\nu_\ell}+\mu_{\nu_\ell}) \right]
}
\\ 
\delta\big((E_\pi -\mu_\pi)-(sE_\ell-\mu_\ell)-(tE_{\nu_\ell}+\mu_{\nu_\ell})\big).
\label{Gamma_pi_gen}
\end{multline}
A similar argument shows that 
\begin{equation}
\Gamma_{\pi^+}(\mu_\pi,\mu_\ell,\mu_{\nu_\ell})
=\Gamma_{\pi^-}(-\mu_\pi,-\mu_\ell,-\mu_{\nu_\ell}).
\end{equation}

\subsection{Pion decay rate at chemical equilibrium}

We are interested in finding a configuration where pions and leptons are in chemical equilibrium, which is expressed by the relation among chemical potentials 
\begin{equation}
\mu_\pi=\mu_\ell-\mu_{\nu_\ell}.
\label{mu-eq}
\end{equation}
Assuming the above relation to be valid, we can solve exactly Eq.~(\ref{Gamma_pi_gen}). 
We proceed in this form, but in Sec.~\ref{pi-l_eq} we explain the necessary conditions for chemical equilibrium. 
Summing up $s$ and $t$ and 
integrating in the neutrino momentum solid angle, considering  $m_\pi>m_\ell$ we obtain
\begin{multline}
\Gamma_{\pi^-}  = \bar\Gamma_{\pi\ell} \frac{m_\pi}{E_\pi}\frac{1}{2a_\ell |{\boldsymbol p}|}
\int_{a_\ell (E_\pi-|{\boldsymbol p}|)}^{a_\ell (E_\pi+|{\boldsymbol p}|)} dE_{\nu_\ell}
\left[1-n_\ell -n_{\bar\nu_\ell}\right]
\label{Gamma_pi_int}
\end{multline}
where $E_\ell = E_\pi-E_{\nu_\ell}$ by energy conservation.
The constant $a_\ell$ is defined as
\begin{equation}
a_\ell\equiv \frac{m_\pi^2-m_\ell^2}{2m_\pi^2} \;,
\end{equation}
while
\begin{equation}
\bar\Gamma_{\pi\ell} =
\frac{1}{\pi}f_\pi^2G_F^2 m_\pi m_\ell^2a_\ell^2
\end{equation}
is the well-known pion decay width in vacuum for $\pi^-\to \ell +\bar\nu_\ell$.

Equation (\ref{Gamma_pi_int}) shows explicitly the particle statistics averaged in the allowed range of energy for the emergent neutrinos, as well as the on-shell energy conservation. 
It is usually interpreted as the probability for inserting a lepton and an antineutrino into the system, minus the probability of the system to produce a lepton and an antineutrino: $(1-n_\ell)(1-n_{\bar \nu_\ell}) -  n_\ell~ n_{\bar \nu_\ell} = 1-n_\ell -n_{\bar \nu_\ell}$. 
In other words, the direct decay rate minus the recombination process $\pi^- \leftrightarrow \ell+\bar\nu_\ell$. 
 So, this rate can be described in terms of the \emph{direct} rate $\Gamma^d_{\pi^-} = \int(1-n_\ell)(1-n_{\bar\nu_\ell})$  and the \emph{inverse} rate $\Gamma^i_{\pi^-}= \int n_\ell \, n_{\bar\nu_\ell}$ as   $\Gamma_{\pi^-}=\Gamma_{\pi^-}^d-\Gamma_{\pi^-}^i$.\footnote{
 Here we adopt the notation of Ref.~\cite{Weldon:1983jn}. 
However, the notation where $ \Gamma^>=\Gamma^d $ and $\Gamma^<=\Gamma^i $ is more standard \cite{lebellac}. }  These rates are related to each other through 
\begin{equation}
\frac{\Gamma_{\pi^-}^d}{\Gamma_{\pi^-}^i} = e^{(E_\pi-\mu_\pi)/T},
\label{Gpi-ratio}
\end{equation}
result that can be verified from Eq. (\ref{Gamma_pi_int}). 
This is an extension of the same result obtained in \cite{lebellac, Weldon:1983jn}, including now chemical potential.\\
From Eq. (\ref{Gpi-ratio}) it can also be obtained 
 \begin{align}
\Gamma_{\pi^-}^d & =e^{\beta(E_\pi-\mu_\pi)}n_{\pi^-}\Gamma_{\pi^-}\\
\Gamma_{\pi^-}^i & = n_{\pi^-}\Gamma_{\pi^-}.
\end{align}

Although this microscopic interpretation can be easily understood (unlike the case of a fermion decaying into a boson and fermion, as we shortly see), it is better to adopt the macroscopic interpretation: $\Gamma_{\pi^-}$ is the rate at which the slightly out-of-equilibrium  pion system can reach thermal equilibrium~\cite{Weldon:1983jn, lebellac},
\begin{equation}
f(t)= n_{\pi-} + c_\pi e^{-t\,\Gamma_{\pi^-}},
\label{f_pi-t}
\end{equation}
where $f(t)$ is the nonequilibrium pion distribution function and $c_\pi$ some energy dependent function.
From now onward we refer to $\Gamma_{\pi^-}$ as the equilibrium deviation decay rate (EDDR).

\bigskip
Integrating Eq.~(\ref{Gamma_pi_int}) with respect to neutrino momentum, we get
\begin{align}
\Gamma_{\pi^-} & = \bar\Gamma_{\pi\ell} \frac{m_\pi}{E_\pi}
\left[1+\frac{T}{2a_\ell |{\boldsymbol p}|}\ln\left(
\frac{1+e^{-(E_\ell^+ - \mu_\ell)/T}}{1+e^{-(E_\ell^- - \mu_\ell)/T}}
\right)\right.
\nonumber\\&\qquad\quad\quad\;
\left.+\frac{T}{2a_\ell |{\boldsymbol p}|}\ln\left(
\frac{1+e^{-(E_{\nu_\ell}^+ + \mu_{\nu_\ell})/T}}{1+e^{-(E_{ \nu_\ell}^- + \mu_{\nu_\ell})/T}}
\right)\right]\;,
\label{Gamma_pi}
\end{align}
where the energy terms are defined as 
\begin{align}
E_\ell^\pm  & =  (1-a_\ell)E_\pi \pm a_\ell |{\boldsymbol p}|  \label{El-pm1},
\\
E_{\nu_\ell}^\pm  & = a_\ell( E_\pi \pm  |{\boldsymbol p}|) \label{Enu-pm1},
\end{align}

\subsection{Pion EDDR in the rest frame}

The  EDDR in the rest frame can be obtained easily even considering the system out of chemical equilibrium.
Let us define a shifted pion mass as
\begin{equation}
\tilde m_\pi \equiv m_\pi-\mu_\pi+\mu_\ell -\mu_{\nu_\ell}\; .
\end{equation}
From Eq. (\ref{Gamma_pi_gen}), setting ${\boldsymbol p}=0$ and considering $m_\ell \leq \tilde m_\pi$, the pion  EDDR is 
\begin{equation}
\Gamma_{\pi^-} =  \bar \Gamma_{\pi\ell}
\frac{\tilde a_\ell^2}{a_\ell^2}
\left[1-n_F( M_\ell-\mu_\ell) - n_F( M_{\nu_\ell} +\mu_{\nu_\ell})\right],
\label{Gamma1}
\end{equation}
where the mass terms  
\begin{align}
M_\ell &\equiv |(1-\tilde a_\ell)\,\tilde m_\pi|,\label{Mell}
\\
M_{\nu_\ell} &\equiv |\tilde a_\ell\,\tilde m_\pi|,
\end{align}
are the rest energy of the terms in Eqs. (\ref{El-pm1}) and (\ref{Enu-pm1}), respectively, but with the pion mass shifted with the chemical asymmetry term, and with the shifted constant 
\begin{equation}
\tilde a_\ell \equiv \frac{\tilde m_\pi^2-m_\ell^2}{2\tilde m_\pi^2}.
\end{equation}
 We observe once again in Eq. (\ref{Gamma1})  the microscopic interpretation in terms of  availability of phase space to allow the resulting leptons to be created: $(1-n_\ell)(1-n_{\bar \nu_\ell}) -  n_\ell~ n_{\bar \nu_\ell} = 1-n_\ell -n_{\bar \nu_\ell}$.

The condition needed to obtain Eq. (\ref{Gamma1}) in this out-of-equilibrium system can be expressed in terms of the neutrino chemical potential.
 The condition $m_\ell\leq \tilde m_\pi$ implies that $\mu_{\nu_\ell}\leq \mu_{\nu_\ell}^c$, where 
 \begin{equation}
\mu_{\nu_\ell}^c \equiv (m_\pi-\mu_\pi)+(\mu_\ell-m_\ell).
 \label{munuc}
 \end{equation}
 This critical quantity is positive if we consider the conditions imposed in Sec.~\ref{considerations}: that charged pions are in normal phase ($\mu_\pi<m_\pi$) and charged leptons are degenerated ($\mu_\ell>m_\ell$). The condition in Eq.~(\ref{munuc}) is valid also for negative values of the neutrino chemical potential, where the thermal bath contains more antineutrinos than neutrinos.

\subsection{High neutrino density}

It is interesting to see what happens if neutrino chemical potential increases beyond $\mu^c_{\nu_\ell}$.

\begin{itemize}

\item If $0\leq\tilde m_\pi \leq m_\ell$, or equivalently 
$0\leq\mu_{\nu_\ell}-\mu_{\nu_\ell}^c\leq m_\ell$, the annihilation of pions induced by scattering with neutrinos from the thermal bath  is  favored: $\pi^-+\nu_\ell \leftrightarrow \ell$.
The EDDR in this case is 
\begin{equation}
\Gamma_{\pi^-} =  \bar\Gamma_{\pi\ell}
\frac{\tilde a_\ell^2}{a_\ell^2}
\left[n_F( M_{\nu_\ell} -\mu_{\nu_\ell})-n_F( M_\ell-\mu_\ell) \right].
\label{Gamma2}
\end{equation}
Since $n_{\nu_\ell}-n_{\ell} = n_{\nu_\ell}(1-n_{\ell})-n_{\ell}(1-n_{\nu_\ell})$, the previous result can be interpreted as the probability of finding a neutrino in the thermal bath from which the pion will scatter off, minus the inverse process.

\item If $-m_\ell \leq \tilde m_\pi \leq 0$, or equivalently  
$m_\ell\leq\mu_{\nu_\ell}-\mu_{\nu_\ell}^c \leq 2m_\ell$, it is more favorable that pions scatter off from antileptons generating neutrinos: $\pi^-+\bar\ell \leftrightarrow \nu$.
From energy conservation, the EDDR is
\begin{equation}
\Gamma_{\pi^-} =  \bar\Gamma_{\pi\ell}
\frac{\tilde a_\ell^2}{a_\ell^2}
\left[ n_F( M_{\nu_\ell} -\mu_{\nu_\ell}) - n_F( M_\ell+\mu_\ell) \right]\;.
\label{Gamma3}
\end{equation}
Notice that $n_{\bar\ell}-n_{\nu_\ell} =n_{\bar\ell}(1-n_{\nu_\ell})-n_{\nu_\ell}(1-n_{\bar\ell})$, which gives the probability of finding an antilepton in the thermal bath from which the pion scatters off leaving a neutrino, minus the probability of the inverse process.

\item Finally, if $\tilde m_\pi \leq -m_\ell$, or equivalently 
$ 2m_\ell\leq\mu_{\nu_\ell} -\mu_{\nu_\ell}^c $,
a $\pi^-$ is totally annihilated by an  antilepton and a neutrino, disappearing in the thermal bath. The opposite reaction means that the thermal bath produces those particles,
$\pi^-+\bar\ell+\nu \leftrightarrow \textrm{(thermal~bath)}$.
Thus,  EDDR in this case is 
\begin{equation}
\Gamma_{\pi^-} =  \bar\Gamma_{\pi\ell}
\frac{\tilde a_\ell^2}{a_\ell^2}
\left[n_F( M_\ell+\mu_\ell) + n_F( M_{\nu_\ell} -\mu_{\nu_\ell})-1\right]\;.
\label{Gamma4}
\end{equation}
Writing $n_{\bar\ell}+n_{\nu_\ell}-1 = n_{\bar\ell}\,n_{\nu_\ell}-(1-n_{\bar\ell})(1-n_{\nu_\ell})$, we interpret this rate as the probability to find an antilepton and a neutrino in the thermal bath minus the probability to introduce an antilepton and a neutrino into the thermal bath.
Backreaction means the thermal bath can spontaneously generate a pion, an antilepton and a neutrino.
\end{itemize}

All the above-mentioned EDDRs are positive. 
These reactions are possible because we are out of chemical equilibrium in a high neutrino density environment.
The meaning of those processes is explained in \cite{Weldon:1983jn} at zero chemical potentials. 
However, we recall that the microscopical description  is not well understood, especially when dealing with the  fermion EDDR. 
The advantage of calculating the imaginary part of the semileptonic self-energy instead of using the Fermi golden rule is because in the latter,  thermal  probability factors must be added by hand.

\section{Metastability conditions}\label{metastability}

For metastability, we understand a special situation where the decay rate of a certain particle becomes much smaller (a bigger lifetime) than the usual case, in analogy with what happens in nuclear isomerism.
In our case, this special situation is produced by the  environment that inhibits the particle  decay and is related to the Pauli blocking effect.
Only thermal fluctuations are capable of overcoming this situation. 
In order to describe this scenario, let us separate the  direct decay rate into the zero temperature and thermal fluctuation parts
\begin{equation}
\Gamma_{\pi^-}^{d} = \Gamma_0 +\delta \Gamma_T
\end{equation}
where the first term is defined as $\Gamma_0 \equiv \lim_{T\to 0}\Gamma_{\pi^-}$ and the second term is the thermal fluctuation.
The temperature for our analysis must be higher than the pion condensation critical temperature.
With this definition we say then that the particle is metastable if $\Gamma_0 =0$.

Let us analyze the case where the system is in chemical equilibrium.
From Eq. (\ref{Gamma_pi}), we have 
\begin{align}
\Gamma_0 = \bar\Gamma_{\pi\ell}\frac{m_\pi}{E_\pi}\Bigg[ 1 
&+ \frac{\mu_\ell-E_\ell^+}{2a_\ell |\bs{p}|}\theta(\mu_\ell-E_\ell^+)
\nonumber \\ &
-\frac{\mu_\ell-E_\ell^-}{2a_\ell |\bs{p}|}\theta(\mu_\ell-E_\ell^-)
\Bigg].
\end{align}
We can immediately see that if $\mu_\ell > E_\ell^+$, the zero temperature part of the pion decay rate vanishes.
This condition depends on the pion energy and the lepton chemical potential. 
Therefore, in order to get $\Gamma_0=0$, the pion energy must be less than a certain critical energy
\begin{equation}
E_\pi<E_c\equiv (1+b_\ell)\,\mu_\ell - b_\ell\, q_F\;,
\end{equation}
or, in terms of the pion momentum,
\begin{equation}
|{\boldsymbol p}| < p_c \equiv (1+b_\ell)\, q_F - b_\ell\, \mu_\ell ,
\label{p_c}
\end{equation}
with
\begin{equation}
b_\ell \equiv \frac{m_\pi^2-m_\ell^2}{2m_\ell^2}\;,
\end{equation}
and where
\begin{equation}
q_F=\sqrt{\mu_\ell^2-m_\ell^2}
\end{equation}
is the lepton Fermi momentum.
From the critical values exposed, one can see that the leptonic chemical potential is restricted to be higher than a certain critical value
\begin{equation}
\mu_\ell  >  \mu_c \equiv (1-a_\ell)m_\pi.
\label{mu_crit}
\end{equation}
At chemical equilibrium, considering muons as the resulting leptons with $m_\pi=139.5\,$MeV, we have the critical value $\mu_c\approx 109.74$~MeV.

\bigskip

If we analyze the hypothetical system with $\pi^+$,  metastability  can be reached if the neutrino chemical potential is high enough so that $\mu_{\nu_\mu}-\mu_{\mu} \geq a_\ell m_\pi$.
This was noticed in \cite{Abuki:2009hx} but at zero temperature, where  pions should condense.

\section{Pion-Lepton thermal equilibrium}
\label{pi-l_eq}

The system we are interested in analyzing is in chemical equilibrium, but not in thermal equilibrium, with metastable pions decaying slowly.
The EDDR in a thermalized system must be seen, as it is explained in Ref.~\cite{Weldon:1983jn}, in terms of the time needed for a system to reach thermal equilibrium [see Eq. (\ref{f_pi-t})].
The decay of pions and the inverse process consider, statistically speaking, pions as real test particles and  ``virtual'' leptons and neutrinos from the heath bath.
We also have to consider another process involving the same participants~\cite{Kuznetsova:2008jt, Kuznetsova:2010pi} if we want to explore the system as a whole: real test leptons recombined with antineutrinos from the heat bath increasing the pion population. 
The latter process occurs very slowly since it is a purely thermal fluctuating phenomenon. 
To compare how slow  metastable pions can decay, we explore under which conditions pions and leptons reach thermal equilibrium simultaneously.
This obviously happens if both EDDR are of the same magnitude: $\Gamma_\ell \sim \Gamma_{\pi^-}$.

%
%
%
%

\bigskip

 \begin{figure}
\includegraphics{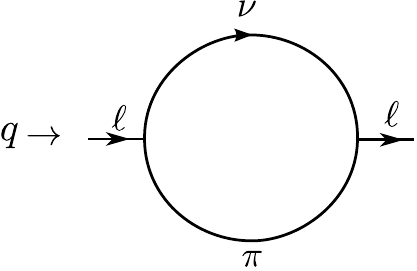}
\caption{Feynman diagram representing the lepton self-energy from weak interactions in a thermal bath.}
\label{leptonSE}
\end{figure}

We need then to find the lepton EDDR assuming  the system is in chemical equilibrium, i.e., $\mu_\pi=\mu_\ell-\mu_{\nu_\ell}$.
The dressed lepton propagator  is 
\begin{equation}
S_\ell^{\textrm{dr}}(q) = \frac{i}{\slashed{Q}-m-\Sigma}\;,
\label{Sdress}
\end{equation}
where $Q=(q_0+\mu_\ell,\bs{q})$ and $\Sigma(q)$ is the lepton self-energy generated by weak coupling, shown in the Feynman diagram in Fig.~\ref{leptonSE}, which corresponds to
\begin{equation}
-i\Sigma(q) = -\int\frac{d^4 p}{(2\pi)^4}V(p)S_{\nu_\ell}(k)V(p)D_{\pi^-}(p)\;,
\end{equation}
with $p=q-k$ by virtue of energy conservation and $V(p)$ as already defined in Eq. (\ref{vertex}).
The general structure of the self-energy in this process is  $\Sigma=\slashed{A}(1-\gamma_5)$, where
\begin{equation}
A_\mu(q) = G_F^2f_\pi^2\int\frac{d^4 k}{(2\pi)^4}\frac{P^2K_\mu-2K\punto P P_\mu}{(P^2-m_\pi^2)K^2}\;.
\end{equation}
Here, $K=(k_0+\mu_{\nu_\ell},\bs{k})$ and $P=(p_0+\mu_\pi,\bs{p})$.
Using the general structure of the self-energy, and considering that, because of spatial symmetry, ${\boldsymbol A}=\hat{\boldsymbol  q}|{\boldsymbol A}|$, the dressed lepton propagator can be written as
\begin{equation}
S_\ell^{\textrm{dr}}=(\slashed{Q}+m-\Sigma)\left[
\frac{\cal P_+}{Q^2-m_\ell^2-\Pi_+}
+
\frac{\cal P_-}{Q^2-m_\ell^2-\Pi_-}
\right],
\end{equation}
where the helicity projectors are 
\begin{equation}
{\cal P}_\pm =\frac{1}{2}(1\pm\gamma_0{\boldsymbol \gamma}\punto \hat{\boldsymbol  q}\gamma_5)\;,
\end{equation}
and the mass corrections are 
\begin{equation}
\Pi_\pm(q) =2Q\cdot A\pm (Q_0{\boldsymbol A}\punto\hat{\boldsymbol  q}-|{\boldsymbol q}|A_0).
\end{equation}
Now the procedure is the same as in Sec.~\ref{decay_width} by setting  $q_0\to q_0+i\epsilon$ in the retarded propagator. 
The EDDR for each lepton helicity is
\begin{equation}
\Gamma_{\pm} = -\frac{1}{E_\ell}\mathrm{Im} ~\Pi_\pm(E_\ell-\mu_\ell+i\epsilon,~{\boldsymbol q}).
\end{equation}
In order to simplify the analysis, we  calculate the average of the EDDRs for the different helicities. 
Defining $\Gamma_\ell = (\Gamma_++\Gamma_-)/2$, after Matsubara summation, the average lepton EDDR is
\begin{align}
\Gamma_{\ell} & = \bar\Gamma_{\pi\ell}\left(\frac{m_\pi}{2m_\ell}\right)^3
\frac{m_\ell}{E_\ell}\frac{1}{2b_\ell |{\boldsymbol q}|}
\int_{b_\ell(E_\ell-|{\boldsymbol q}|)}^{b_\ell(E_\ell+|{\boldsymbol q}|)} dE_{\nu_\ell}
\left[n_{\pi^-}+n_{\bar\nu_\ell}\right]
\label{Gamma_ell_int}
\end{align}
where $E_\pi = E_\ell +E_{\nu_\ell}$ by energy conservation.
 
The lepton EDDR is related to a direct process, where an external lepton scatters with a thermalized antineutrino, giving rise to a thermalized pion $\ell+\bar\nu_\ell\to \pi$. 
The corresponding inverse process can be thought as a thermalized pion decaying into a thermalized antineutrino and a observable lepton $\pi\to \ell + \bar\nu$.
In terms of rates, $\Gamma_\ell =\Gamma_\ell^d +\Gamma_\ell^i = \int(n_{\pi^-}+n_{\bar\nu_\ell})$, where the direct rate is $\Gamma_\ell^d = \int(1+n_{\pi^-})n_{\bar \nu_\ell}$ and the inverse rate is $\Gamma_\ell^i  = \int n_{\pi^-}(1-n_{\bar\nu_\ell})$.
Here, the direct  and the inverse processes have to be added in opposition to the case of boson decay.
This is a general feature of decaying fermions as was pointed in \cite{Weldon:1983jn}.

The direct and inverse rates are related as
\begin{equation}
\frac{\Gamma_\ell^d}{\Gamma_\ell^i} = e^{(E_\ell-\mu_\ell)/T},
\end{equation}
and also from this result, these rates can be written as
\begin{align}
\Gamma_\ell^d &= e^{(E_\ell-\mu_\ell)/T}n_\ell \Gamma_\ell\;, \\
\Gamma_\ell^i &= n_\ell \Gamma_\ell \;.
\end{align}

The lepton EDDR is interpreted as the rate at which the slightly out-of-equilibrium lepton system can reach the thermal equilibrium:
\begin{equation}
f(t) = n_\ell+ c_\ell e^{-t\, \Gamma_\ell}
\end{equation}
where $f(t)$ is the nonequilibrium lepton distribution function and $c_\ell$ some energy dependent function.

\bigskip

Integrating out the neutrino momentum, we obtain the general result
\begin{align}
\Gamma_{\ell}  = \bar\Gamma_{\pi\ell} \left(\frac{m_\pi}{2m_\ell}\right)^3
\frac{m_\ell}{E_\ell}
&
\frac{T}{2b_\ell |{\boldsymbol q}|}
\Bigg[ 
\ln\left(\frac{1-e^{-(E_\pi^+ -\mu_\pi)/T}}{1-e^{-(E_\pi^- -\mu_\pi)/T}}\right)\;,
\nonumber \\ &
-\ln\left(
\frac{1+e^{-( \tilde E_{\nu_\ell}^+ +\mu_{\nu_\ell})/T}} {1+e^{-( \tilde E_{\nu_\ell}^- +\mu_{\nu_\ell})/T}}\right) 
\Bigg] \;,
\label{Gamma_ell}
\end{align}
where 
\begin{align}
    E_\pi^\pm  & =  (1+b_\ell)E_\ell \pm b_\ell |{\boldsymbol q}| \;, \label{El-pm}\\
 \tilde E_{\nu_\ell}^\pm  & = b_\ell E_\pi \pm b_\ell  |{\boldsymbol q}| \label{Enu-pm}.
\end{align}

\bigskip

\begin{figure}
\includegraphics[width=\columnwidth]{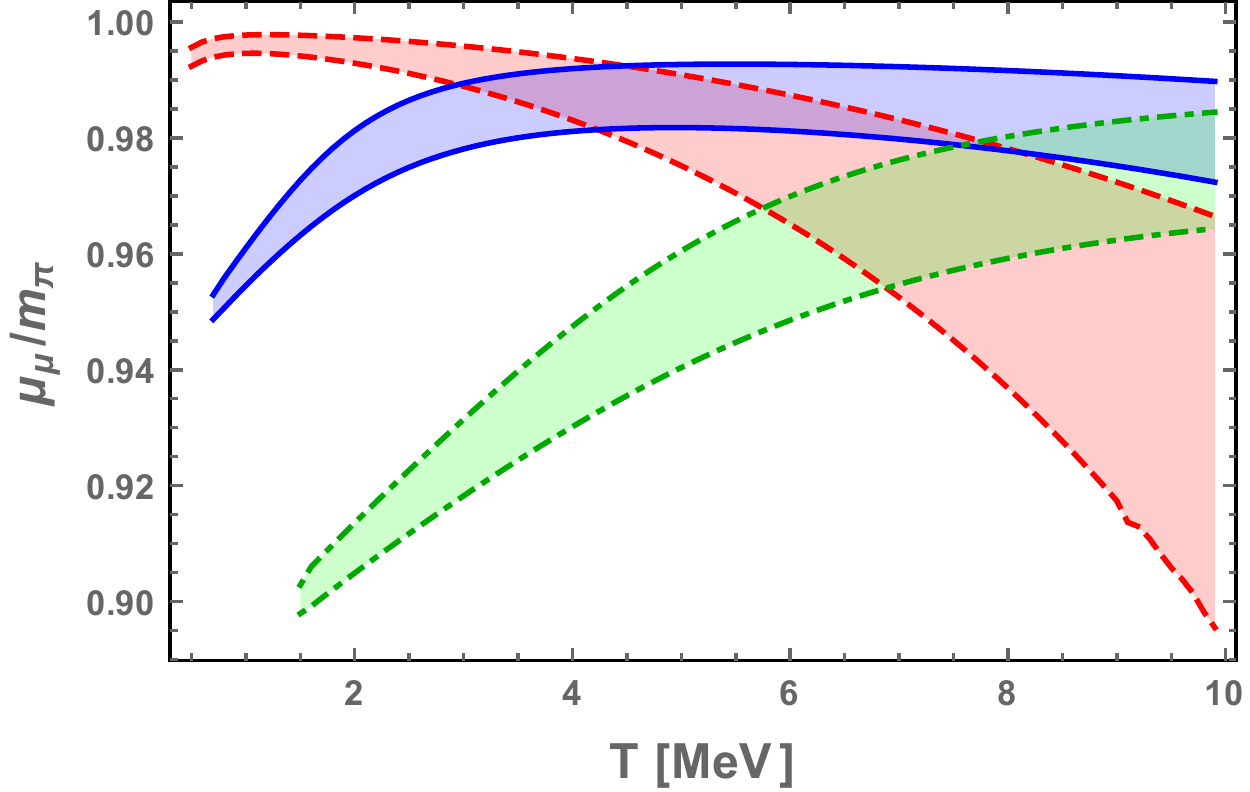}\\
\vspace{.5cm}
\includegraphics[width=\columnwidth]{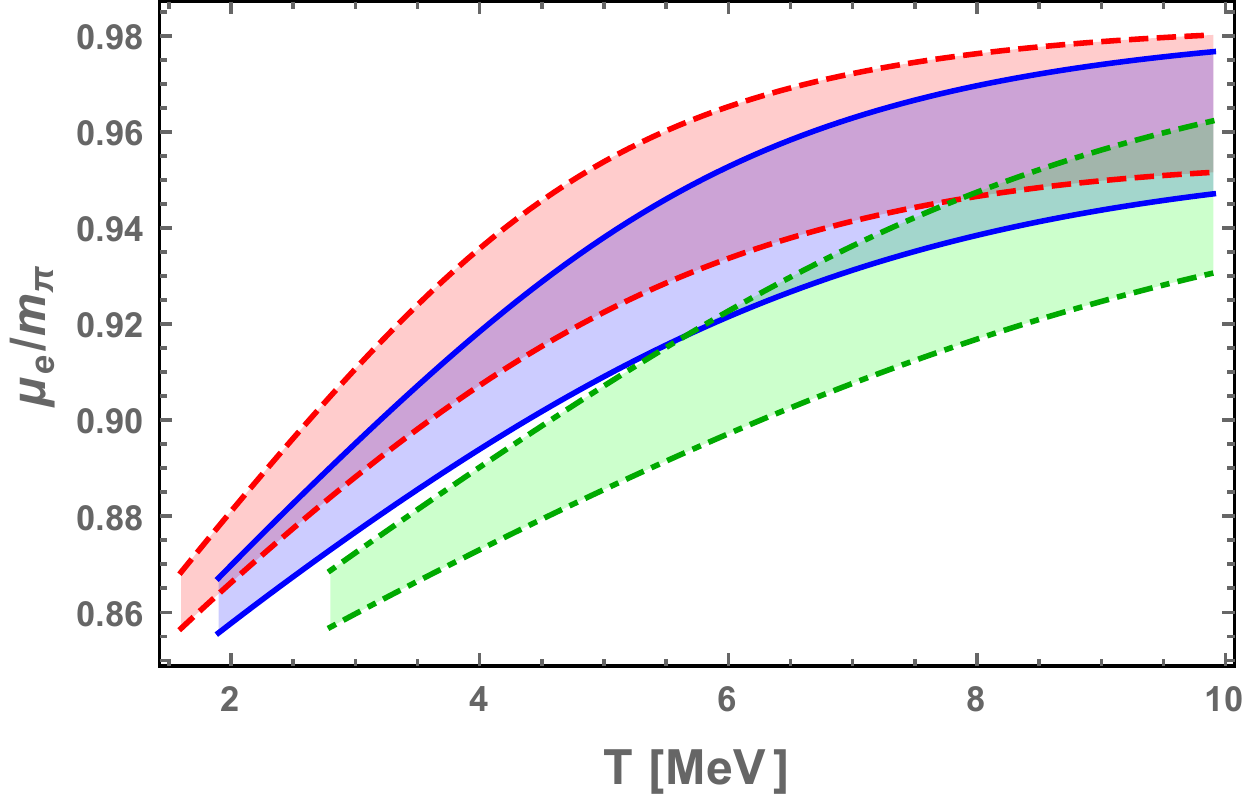}
\caption{Chemical potential and temperature values where pion-lepton chemical equilibrium is favorable. The band widths correspond to a half order of magnitude difference between the widths: $0.5 < \Gamma_{\pi^-}/\Gamma_{\ell} < 1.5$.
The colors and lines refer to $m_\pi =140$~MeV (solid blue),  $m_\pi =115$~MeV (dashed red) and $m_\pi =200$~MeV (dot-dashed green).}
\label{mueq}
\end{figure}

Now, we need to find a window in the parameter space where pions and leptons reach thermal equilibrium satisfying $\Gamma_\ell\sim\Gamma_{\pi^-}$.
The parameters involved are the neutrino chemical potential, the lepton chemical potential, the temperature, the pion energy and the lepton energy. 
To reduce the number of parameters we consider pions  in the rest frame, where the decay rate is given in Eq. (\ref{Gamma1}), with $\tilde m_\pi \to m_\pi$.
In the same way, near the Fermi surface, fluctuations of leptons in the degenerated environment are produced.
So we take the lepton energy as the Fermi energy in the decay rate, which leads to
\begin{multline}
\Gamma_{\ell}  = \bar\Gamma_{\pi\ell} \left(\frac{m_\pi}{2m_\ell}\right)^3
\frac{m_\ell}{\mu_\ell}
\frac{T}{2b_\ell\, q_F}
\\ \times
\ln\left(\frac{\sinh [(b_\ell\mu_\ell+\mu_{\nu_\ell})/T]+\sinh[b_\ell q_F/T]}{\sinh [(b_\ell\mu_\ell+\mu_{\nu_\ell})/T]-\sinh[b_\ell q_F/T]}\right).
\end{multline}
For simplicity we  consider $\mu_{\nu_\ell}=0$, which means that all real neutrinos escape from the star once created. 
This is a usual approximation in neutron stars, but not necessarily valid for protoneutron stars \cite{Pons:1998mm, Reddy:1996tw, Prakash:1996xs, Pons:2000xf}.
With such considerations, the EDDRs depend only on temperature and lepton chemical potential. 
The regions where $\Gamma_{\pi^-}\sim\Gamma_\ell$ are plotted in Fig.~\ref{mueq}, describing the temperature and lepton chemical potential conditions for reaching thermal equilibrium simultaneously. 
This is considered for pions decaying into muons (upper panel) as well as pions decaying into electrons (lower panel) for different values of $m_\pi$. 
For decay rates such that $0.5 < \Gamma_{\pi^-}/\Gamma_{\ell} < 1.5$, we can see in Fig.~\ref{mueq} that there is a wide region where simultaneous  equilibrium is possible, especially at high temperature, and always for values of the lepton chemical potential greater than the critical chemical potential for metastability, defined in Eq.~(\ref{mu_crit}). 
If pions are condensed they still can decay \cite{Mammarella:2015pxa}; therefore, at finite temperature, such a kind of equilibrium analysis must be considered in the condensed case.

Notice that in compact stars we have leptonic  equilibrium: $\mu_\mu-\mu_{\nu_\mu} = \mu_e-\mu_{\nu_e}$, and also beta equilibrium, $\mu_n-\mu_p = \mu_e-\mu_{\nu_e}$. 
Therefore, if there is pion-lepton equilibrium, then $\mu_n-\mu_p = \mu_\pi$.
 The processes  $n\leftrightarrow p +\pi^-$ are always present since pions strongly interact with nuclear medium.


\section{Neutrino emission}\label{nu_emission}

The cooling process in compact stars  is mainly produced by neutrino emission and provides information about the existence of nontrivial hadronic matter states.
 In particular, neutrino emission through pion decay has been studied in different models \cite{Jaikumar:2002vg,Reddy:2002xc, Schafer:2010cs, Alikhanov:2012cp}
The quantity that governs such a process is the emissivity (energy loss by neutrino emission per unit of time), the Urca process being the most efficient one 
\cite{Yakovlev:2000jp}.
In this section we calculate the neutrino emissivity due to pion decay in a pion-lepton equilibrium regime.

\begin{figure}
\includegraphics[scale=2]{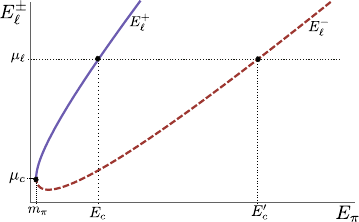}
\caption{ Integration limits of the lepton energy as a function of pion energy. 
The critical chemical potential where metastability starts is indicated, as well as the pion energy values where $E_\ell^\pm = \mu_\ell$.}
\label{El-pm_Epi}
\end{figure}

The neutrino emissivity generated by leptonic decay of pions is defined as the transition
probability matrix multiplied by: the ejected neutrino energy,  the probability $n_\pi^-$ of finding a pion in the thermal bath, and  the probability $1-n_\ell$ of finding a hole below the Fermi level. All this integrated in phase space. Therefore
\begin{multline}
\epsilon_{\pi} =  
\int \bar{d}p\,\bar{d}q\,\bar{d}k\,
\sum_{\text{spin}}|{\cal M}|^2\, k_0\, n_B(p_0)\, [1-n_F(q_0)]\\
(2\pi)^4\delta^4(p-q-k),
\end{multline}
where the phase space measure for pions is
\begin{equation}
\bar dp = \frac{d^4 p}{(2\pi)^3}\theta(p_0+\mu_\pi)\delta^4((p_0+\mu_\pi)^2-E_\pi^2)\;,
\end{equation}
and an equivalent expression for the leptons.
The probability amplitude for pions going into leptons and antineutrinos is defined as 
\begin{equation}
\langle \ell\,\bar\nu| \!\int\!\! d^4x\, {\cal L}_{\pi\ell}\, |\pi^-\rangle
=i{\cal M}(2\pi)^4\delta^4(p-q-k).
\end{equation}
At chemical equilibrium, the neutrino emissivity from pions is then
\begin{multline}
\epsilon_{\pi} = \frac{\bar\Gamma_{\pi\ell}\, m_\pi T^2}{2\pi^2 a_\ell}
\int_{m_\pi}^\infty dE_\pi \, n_B(E_\pi-\mu_\pi) 
\\
\bigg[
\frac{E_\pi-E_\ell}{T}\,\ln\!\left(1+e^{(E_\ell-\mu_\ell)/T}\right)
\\
\left.
-{\rm Li}_2\!\left(-e^{(E_\ell-\mu_\ell)/T}\right)
\bigg]\right|_{E_\ell^-}^{E_\ell^+},
\label{emissivity_full}
\end{multline}
where $E_\ell^\pm$ was defined in Eq. (\ref{El-pm}) and plotted in Fig.~\ref{El-pm_Epi} as a function of pion energy.

\begin{figure}
\includegraphics[scale=0.65]{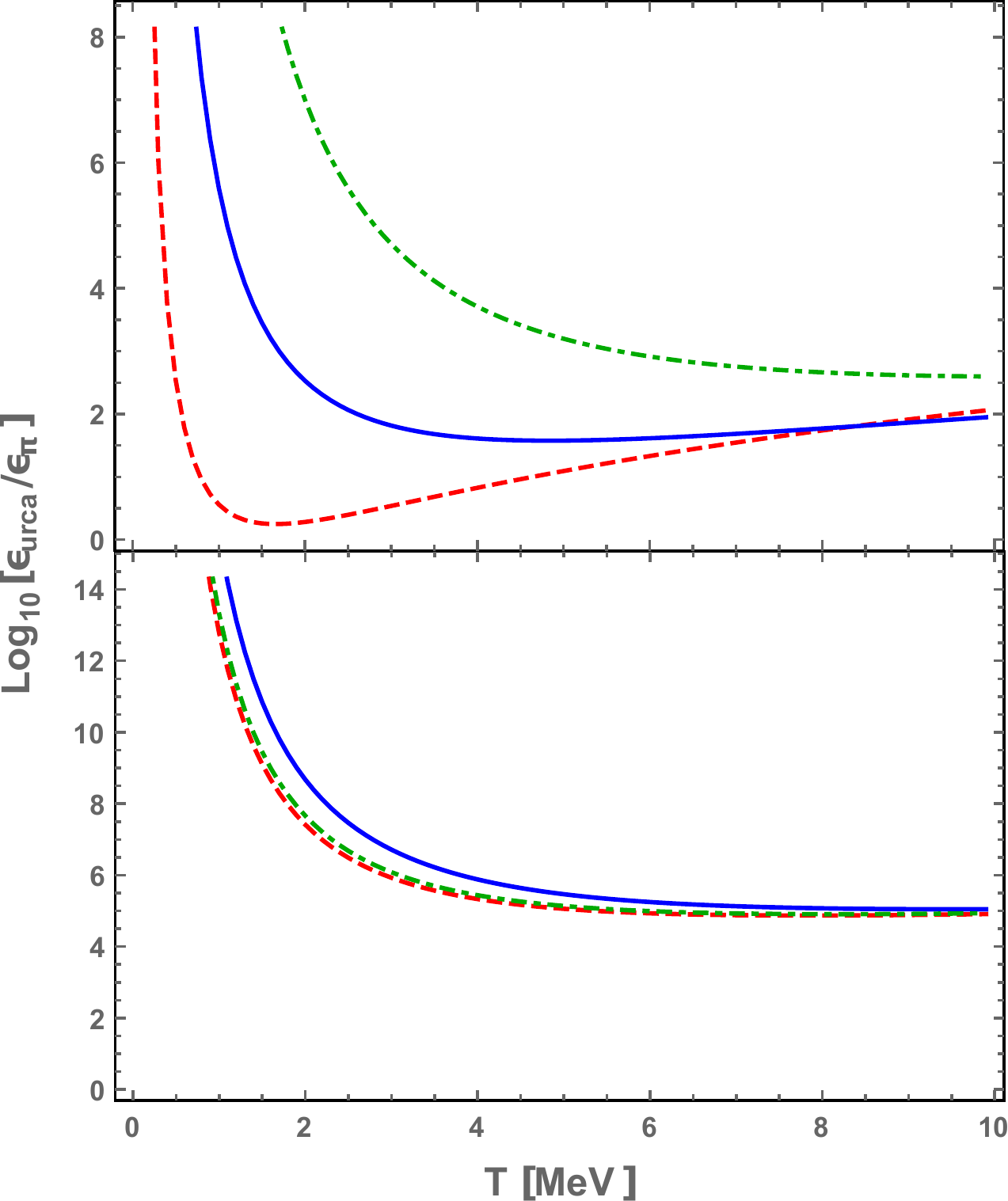}
\caption{Comparison between emissivities produced by the URCA process and pion decaying into muons (upper panel) and electrons (lower panel). 
In both cases we use the values $\mu_\ell=0.95 m_\pi$. $f_\pi = 0.8\times 93.4$ MeV and the pion masses $m_\pi$ =115 (dashed red), 140 (solid blue), and 200 MeV (dot-dashed green). }
\label{emissivity}
\end{figure}

A comparison between the neutrino emissivity due to pion decay and by the URCA process is shown in Fig.~\ref{emissivity} for pion decaying into muons (upper panel) and electrons (lower panel).

URCA emissivity \cite{Yakovlev:2000jp} is given by 
\begin{equation}
\epsilon_{\tiny\textrm{URCA}}= \frac{457\pi}{10080}G_F^2 |V_{ud}|^2
(1+3g_A^2) m_n^* m_p^* m_e^* T^6 \Theta_{npe}\;,\label{URCA}
\end{equation}
where $g_A= 1.2$ is the axial coupling, $m_n^*\approx m_p^* \approx
0.8\times 940~{\rm MeV}$ are the nucleon effective masses (in the Brown-Rho scaling) and $m_e^* \approx \mu_e$
is the electron effective mass. The function
$\Theta_{npe}=\theta(p_{Fp}+p_{Fe}-p_{Fn})$
is the triangular condition between the Fermi momenta of the proton,
electron, and neutron, which we consider here as satisfied.

We see that neutrino emission  tends to be $\sim 10^{-2}$ times lower than the URCA process at high temperature for pions decaying into muons, and $\sim 10^{-5}$ in the case when pions decay into electrons. 
Notice that for a small pion mass, the muonic neutrino emissivity is similar to the URCA emissivity at $T\sim 1$~MeV.

\bigskip

At temperature much higher than 20~MeV, the star turns to be opaque to neutrinos \cite{Reddy:1996tw}. 
Neutrino total mean free path is inversely proportional to the probability of absorption and scattering with the participants.
Because the metastable state of pions emerges by the difficulty for these particles to decay into degenerated leptons (lepton Fermi levels are almost filled), one can straightforwardly infer that neutrino absorption by pions is highly suppressed in degenerated lepton environments ($\pi^-+\nu_\ell\to\ell$). On the other hand, such a metastable state enlarges the amount of pions in the environment, and thus, the possibility of pion-neutrino scattering ($\pi^-+\nu_\ell\to \pi^-+\nu_\ell$) increases. 
It is not clear which effect dominates. 
The detailed calculation must be done to see the overall effect.

\subsection{Low temperature approximation}

Now we explore how the emissivity behaves in the low temperature approximation. 
The scale here is the pion mass, so $T\sim 10$~MeV is small enough.
To expand the exponentials in Eq.~(\ref{emissivity_full}) we need a negative exponent.
The lepton energy integration limits $E_\ell^\pm(E_\pi)$ are plotted in Fig. \ref{El-pm_Epi}, where all the terms in the plot were already defined except $E'_c=(1+b_\ell)\mu_\ell + b_\ell q_F$.
The procedure for the expansion is to separate the pion energy integral in the ranges where $E_\ell^\pm -\mu$ is negative expanding the exponential in the logarithm and the polylogarithm.
In the integration ranges where   $E_\ell^\pm -\mu$  is positive, the log and polylog can be expanded after the use of the inversion formulas
\begin{align}
\ln(1+e^x)-\ln(1+e^{-x})-x &=  0,\\
{\rm Li}_2(-e^x)+{\rm Li}_2(-e^{-x})+\frac{x^2}{2}+\frac{\pi^2}{6} &= 0.
\end{align}
The Bose-Einstein term is expanded also up to the leading order. 
Keeping the leading terms, the integration in pion energy can be performed exactly from $m_\pi$ to $E_c$ or $E_c'$. 
From $E_c$ and $E_c'$ to infinity, the integrands are expanded around the lowest value due to exponential suppression.
As a result, the emissivity is quite different from the one expected in a simple boson gas, where $\epsilon \sim T^{3/2} e^{-(m_\pi-\mu_\pi)/T}$ \cite{Jaikumar:2002vg,Reddy:2002xc, Schafer:2010cs, Alikhanov:2012cp}. 

When $p_c\gg T$, which implies that lepton chemical potential should be bigger than the critical chemical potential, the emissivity is proportional to $T^2$, namely
\begin{equation}
\epsilon_\pi \approx
\bar\Gamma_{\pi\ell} \,m_\pi^4\,
g(\mu_\ell)\,\left(\frac{T}{2\pi m_\pi}\right)^{2}
 e^{-(E_c-\mu_\pi)/T},
\end{equation}
with
\begin{multline}
g(\mu_\ell)=
\frac{2q_F}{b_\ell m_\pi}+\frac{2a_\ell m_\pi}{p_c}+\frac{2p_c}{a_\ell m_\pi}\frac{q_F-p_c}{q_F+p_c}\;.
\end{multline}

For $p_c \ll T$, which means $\mu_\ell \sim \mu_c$, the emissivity is
\begin{multline}
\epsilon_\pi \approx \left\{
2(1-a_\ell-a_\ell^2)\left(\frac{T}{2\pi m_\pi}\right)^{2} +a_\ell\left( \frac{T}{2\pi m_\pi}\right)^{3/2}
\right\} \\ \times
\bar\Gamma_{\pi\ell} \,m_\pi^4\,e^{-(E_c-\mu_\pi)/T}\;.
\end{multline}
These approximations fairly fit the full expression and explicitly show that the neutrino emission from chemically equilibrated pions is much smaller than it was expected.

\section{Conclusions}\label{conclusions}

In this work we have studied the effects of dense lepton matter over the decay rate of charged pions.
We use  conditions like in the neutron star environment, consisting in degenerated lepton matter and modified hadronic parameters due to dense nuclear matter.
We found that, at certain values of the pion chemical potential, pions drift into a metastable state and their decay is generated only through thermal fluctuations. 
This scenario is generated by the fact that all leptonic energy states are occupied up to the Fermi level, so the pion has no phase space  to decay.
In this way, the pion EDDR is calculated at finite temperature and chemical potentials. 
The particular case of pion EDDR at chemical equilibrium with leptons, as well as the case out of equilibrium in rest frame was studied.

The pion momentum is restricted for momentum below a certain critical momentum, and the allowed values of the chemical potential must be lower than the pion mass and higher than a critical chemical potential. 
These critical parameters are obtained by studying the behavior of the direct pion decay rate at zero temperature.
The thermal equilibrium was also studied in terms of equivalent pion EDDR and lepton EDDR. 
The considered values of temperature and chemical potential for rest frame pions and leptons in the Fermi surface are reasonable.
Finally, we calculate the neutrino emissivity generated from metastable pion decay in chemical equilibrium, and compare with the URCA emissivity. 
The order of magnitude turns closer for $T\sim 4$~MeV and is considerably smaller than the usual boson gas emissivity.

We conclude that this state of matter is more favorable in protoneutron stars where temperature reaches values higher than 1 MeV.
This scenario therefore can have significant repercussions only for a short time. 
In cold compact stars, metastable pions can also be founded,  but are much less abundant. \par
The assumptions we consider can be reproduced for other kind of particles.
In particular, the case of kaons could be a better example in compact stars since their mass reduces considerably due to dense nuclear environment and the possibility to get metastable kaons in cold dense matter in principle is enhanced.

\section*{Acknowledgments}

This work has been supported by FONDECYT (Chile) Grant No. 1130056, No. 1150471, No.   1150847 and No. 1170107, ConicytPIA/Basal (Chile) Grant No. FB0821, CIC-UMSNH (M\'exico)  Grant No. 4.22, and Consejo Nacional de Ciencia y Tecnología (M\'exico) Grant No. 256494.  
CV acknowledges the group {\em F\'isica de Altas Energ\'ias} at UBB.

\end{document}